\documentclass{aa}
\usepackage[varg]{txfonts}
\usepackage{graphics}
\usepackage{graphicx}
\usepackage{color}

\usepackage{epsfig}
\usepackage{array}
\usepackage{ctable}
\usepackage{amssymb}
\usepackage{amsmath}
\usepackage{url}
\usepackage{lscape}
\usepackage{amsfonts}
\usepackage{mathpazo}
\usepackage{color}
\usepackage{fmtcount}
\usepackage{lipsum}
\usepackage{multirow}
\usepackage{subfig}
\usepackage{natbib}
\usepackage{threeparttable}
\bibpunct{(}{)}{;}{a}{}{,}
\usepackage{enumitem}
\usepackage{booktabs}
\usepackage{gensymb}

\begin{document}

\title{Variation in solar differential rotation and activity in the period
1964 - 2016 determined by the Kanzelh{\"o}he data set}

\newcommand{\angstrom}[1]{\AA}

\author{I. Poljan\v{c}i\'{c} Beljan\inst{1}
 \and R. Jurdana-\v{S}epi\'{c}\inst{1}
  \and T. Jurki\'{c}\inst{1}
   \and R. Braj\v{s}a\inst{2}
    \and  I. Skoki\'{c}\inst{2}
     \and D. Sudar\inst{2}
      \and D. Ru\v{z}djak\inst{2}
       \and D. Hr\v{z}ina\inst{3}
        \and W. P{\"o}tzi\inst{4}
         \and A. Hanslmeier\inst{5}
          \and A. M. Veronig\inst{4,5}}

\authorrunning{I. Poljan\v{c}i\'{c} Beljan et al.}
\titlerunning{Variation in solar differential rotation and activity}

\institute{University of Rijeka, Faculty of Physics, Radmile Matej\v{c}i\'{c} 2, 51000 Rijeka, Croatia
\and Hvar Observatory, Faculty of Geodesy, University of Zagreb, Ka\v{c}i\'{c}eva 26, 10000 Zagreb, Croatia
\and Zagreb Astronomical Observatory, Opati\v{c}ka 22, 10000 Zagreb, Croatia
\and Kanzelh{\"o}he Observatory for Solar and Environmental Research, University of Graz, Kanzelh{\"o}he 19, 9521 Treffen am Ossiacher See, Austria
\and Institute of Physics, University of Graz, Universit{\"a}tsplatz 5, 8010 Graz, Austria}

\date{ Received /Accepted }

\abstract{}
{Theoretical calculations predict an increased equatorial rotation and more pronounced differential rotation (DR) during the minimum of solar magnetic activity. However, the results of observational studies vary, some showing less and some more pronounced DR during the minimum of solar magnetic activity. Our study aims to gain more insight into these discrepancies.}
{We determined the DR parameters $A$ and $B$ (corresponding to the equatorial rotation velocity and the gradient of the solar DR, respectively) by tracing sunspot groups in sunspot drawings of the Kanzelh{\"o}he Observatory for Solar and Environmental Research (KSO; 1964-2008, for solar cycles 20-23) and KSO white-light images (2009-2016, for solar cycle 24). We used different statistical methods and approaches to analyse variations in DR parameters related to the cycle and to the phase of the solar cycle, together with long-term related variations.}
{The comparison of the DR parameters for individual cycles obtained from the KSO and from other sources yield statistically insignificant differences for the years after 1980, meaning that the KSO sunspot group data set is well suited for long-term cycle to cycle studies. The DR parameters $A$ and $B$ show statistically significant periodic variability. The periodicity corresponds to the solar cycle and is correlated with the solar activity. The changes in $A$ related to solar cycle phase are in accordance with previously reported theoretical and experimental results (higher $A$ during solar minimum, lower $A$ during the maximum of activity), while changes in $B$ differ from the theoretical predictions as we observe more negative values of $B$, that is, a more pronounced differential rotation during activity maximum. The main result of this paper for the long-term variations in $A$ is the detection of a phase shift between the activity flip (in the 1970s) and the equatorial rotation velocity flip (in the early 1990s), during which both $A$ and activity show a secular decreasing trend. This indicates that the two quantities are correlated in between 1970 and 1990. Therefore, the theoretical model fails in the phase-shift time period that occurs after the modern Gleissberg maximum, while in the time period thereafter (after the 1990s), theoretical and experimental results are consistent. The long-term variations in $B$ in general yield an anticorrelation of $B$ and activity, as a rise of $B$ is observed during the entire time period (1964-2016) we analysed, during which activity decreased, with the exception of the end of solar cycle 22 and the beginning of solar cycle 23.}
{We study for the first time the variation in solar DR and activity based on 53 years of KSO data. Our results agree well with the results related to the solar cycle phase from corona observations. The disagreement of the observational results for $B$ and theoretical studies may be due to the fact that we analysed the period immediately after the modern Gleissberg maximum, where for the phase-shift period, $A$ versus activity also entails a result that differs from theoretical predictions. Therefore, studies of rotation versus activity with data sets encompassing the Gleissberg extremes should include separate analyses of the parts of the data set in between different flips (e.g. before the activity flip, between the activity and the rotation flip, and after the rotation flip).}


\maketitle

%**************************************************************************************************************************
%******************************** I N T R O D U C T I O N  *******************************************************************

\section{Introduction} \label{Introduction}

Studies of the temporal variations in solar differential rotation (DR) and solar activity are important for understanding how the solar magnetic cycle is generated. Sunspots and sunspot groups have very often been used as tracers to investigate temporal variations in solar rotation using the Greenwich Photoheliographic Results (GPR) \citep{balthwoehl1980,arevaloetal1982,balthvawo1986,1998A&A...332..748P}, the data set from the Solar Observing Optical Network/United States Air Force/National
Oceanic and Atmospheric Administration (SOON/USAF/NOAA, abbreviated SOON/NOAA) \citep{pulktuo1998,javaraiah2003,javaraiahul2006,2013SoPh..287..197J,2016Ap&SS.361..208J}, the Mount Wilson data set \citep{1984ApJ...283..373H,gilmanhow1984,hathawaywi1990}, and data set from the Kanzelh{\"o}he Observatory for Solar and Environmental Research (KSO) \citep{lustig1983}. Long-term studies are necessary to verify the results of temporal changes in DR, as well as to clear ambiguities related to the relation between the rotation and activity.

When the sunspot groups are used as tracers, observational results for the sidereal angular velocity at the solar surface are very often described by the solar DR law,
\begin{equation}
\omega(b) = A+B\sin^2b
,\end{equation}
where $\omega$ is the angular rotation velocity, $b$ is the heliographic latitude, the DR parameter $A$ represents the equatorial rotation velocity, and the DR parameter $B$ (as a negative value) represents the gradient of the differential rotation. In other words, $B$ specifies the degree of rotational nonuniformity, that is, the difference in angular velocity between the poles and the equator.
Theoretical studies based on analytical solutions of the angular momentum transport equation
within the convection zone \citep{2004ApJ...614.1073B,2006MNRAS.373..819L,2007A&A...471.1011L,2014IAUS..302..114B,2016AdSpR..58.1507V} predict that the net transport of angular momentum towards the equator is
less efficient during high solar activity. In other words, strong magnetic fields suppress the equatorward angular momentum transport, resulting in reduced equatorial velocities, that is, in a lower value of the DR parameter $A$. As a consequence, during the activity maximum, the rotation is less differential (the difference in angular velocity between the pole and the equator is smaller), which is represented by a less negative DR parameter $B$. Generally, the first theoretical conclusion for parameter $A$ is in accordance with published experimental results: during activity maximum, lower values of $A$ are observed, and during activity minimum, higher values are reported. The behaviour of parameter $B$ does not coincide with theoretical predictions, however: during activity maximum, sometimes less negative and sometimes more negative values of $B$ are observed \citep{lustig1983,gilmanhow1984,balthvawo1986,guptasiho1999,zucczapp2003,2006SoPh..237..365B,2011A&A...534A..17J,2014SoPh..289..759L}. This implies that further analyses are necessary.

This paper is a continuation of the analysis performed by \citet{2017A&A...606A..72P}, hereafter Paper I. The main conclusion of Paper I was that the KSO sunspot group position data represent a valuable data set with satisfactory accuracy that is suitable for investigating the long-term variabilities in the solar rotation profile. Therefore, we analyse the temporal variation in equatorial rotation velocity ($A$) and gradient of the DR ($B$) using the KSO data on cycle, in-cycle, and on longer temporal scales. We also investigate the relation between the relative sunspot numbers from the World Data Center - Sunspot Index and Long-term Solar Observations, Royal Observatory of Belgium, Brussels (WDC-SILSO) as indices of the solar activity and solar rotation described with the parameters $A$ and $B$ determined from the KSO sunspot group position data (1964 – 2016). The results are then compared with previous works on this topic \citep{2006SoPh..237..365B,2011A&A...534A..17J,2014SoPh..289..759L,2017SoPh..292..179R,2017MNRAS.466.4535B,2020SoPh..295..170J}.

Previous studies of solar drawings from the KSO, which analysed the DR of the sunspots and sunspot groups, were based on data before 1985. The same is true for the analyses of the temporal variation in DR. The only reference that analysed the temporal variation in DR of the KSO data is \citet{lustig1983}, and it covers the period 1947-1981, in which solar cycle (SC) 21 was partly covered. We here extend the analysis to 2016. For the first time, the analysis was performed with 53 years of KSO data, covering five SCs 20-24. The same holds for this analysis of the relation between solar rotation and solar activity. In addition to extending the temporal variation analysis, one of the main goals of the present paper is to gain clearer insight into the discrepancies between observations and theoretical predictions for the relation between $B$ and solar activity.

The measurements and data reduction are described in Sect. 2. The results are presented and discussed in Sect. 3,  which consists of three subsections covering the three main topics of the paper: cycle-related variations in DR (Sect. 3.1.), solar cycle phase variations in DR (Sect. 3.2.), and long-term variations in DR (Sect. 3.3.). The conclusions are drawn in Sect. 4.

%/////////////////////////////// TABLE 1 - references ///////////////////////

\begin{table*}[!ht]
\begin{center}
\caption{Values of DR parameters $A$ and $B$ and their standard errors (in deg/day) for SCs 20 - 24 collected from different sources and determined by the use of sunspots and sunspot groups as tracers.}\label{Tab1}
\begin{tabular}
[c]{>{\centering}m{0.30cm}>{\centering}m{2.4cm}c>{\centering}m{1.7cm}>{\centering}m{2.4cm}ccc}\hline\hline\noalign{\smallskip}

Row&Data set    & Time (Cycle)   &Hemisphere& $A$  & $B$&References     \\\hline\noalign{\smallskip}

1&GPR\tablefootmark{a}      &1965 - 1976 (20) &N+S      &14.53  $\pm$0.02       &-2.73  $\pm$0.17       &1      \\
2&SOON/NOAA\tablefootmark{b}       &1965 - 1977 (20)&N+S        &14.522 $\pm$0.014      &-2.593$\pm$0.138       &2\\
3&      GPR\tablefootmark{a}    &       1965 - 1975 (20)        &       N+S         &       14.502  $\pm$   0.010   &       -2.372  $\pm$   0.114&3\\
4&      GPR\tablefootmark{a}    &       1965 - 1975 (20)        &       N         &       14.492  $\pm$   0.015   &       -2.392  $\pm$   0.149&3\\
5&KSO\tablefootmark{b}  &       1965 - 1976 (20)        &       N       &       14.40   $\pm$   0.02    &       -2.75   $\pm$   0.24&5  \\
6&GPR\tablefootmark{a}  &       1965 - 1975 (20)        &       S       &       14.506  $\pm$   0.015   &       -2.298  $\pm$   0.183&3 \\
7&KSO\tablefootmark{b}  &       1965 - 1976 (20)        &       S       &       14.37   $\pm$   0.02    &       -2.48   $\pm$   0.27&5  \\

\hline\noalign{\smallskip}

8&SOON/NOAA\tablefootmark{b}       &1977 - 1987 (21)&N+S        &14.492 $\pm$   0.020   &-2.696$\pm$    0.172&2\\
9&      SOON/NOAA\tablefootmark{a}      &       1976 - 1986 (21)        &       N+S         &       14.492  $\pm$           0.015   &       -2.080  $\pm$           0.109&3\\
10&     SOON/NOAA\tablefootmark{a}      &       1976 - 1986 (21)        &       N         &       14.477  $\pm$           0.019   &       -1.892  $\pm$           0.158&3 \\
11&KSO\tablefootmark{b} &       1976 - 1981 (21)        &       N       &       14.50   $\pm$   0.02    &       -2.57   $\pm$   0.20&5  \\
12&     SOON/NOAA\tablefootmark{a}      &       1976 - 1986 (21)        &       S         &       14.502  $\pm$           0.019   &       -2.249  $\pm$           0.154&3\\
13&KSO\tablefootmark{b} &       1976 - 1981 (21)        &       S       &       14.47   $\pm$   0.03&   -2.34   $\pm$   0.20&5  \\

\hline\noalign{\smallskip}

14&SOON/NOAA\tablefootmark{b}      &1987 - 1997 (22)&N+S        &14.436 $\pm$   0.018   &-2.286$\pm$    0.146   &2\\
15&     SOON/NOAA\tablefootmark{a}      &       1987 - 1996 (22)        &       N+S         &       14.413  $\pm$           0.015   &       -2.060  $\pm$           0.119&3\\
16&     SOON/NOAA\tablefootmark{a}      &       1987 - 1996 (22)        &       N         &       14.428  $\pm$           0.019   &       -2.011  $\pm$           0.168&3 \\
17&     SOON/NOAA\tablefootmark{a}      &       1987 - 1996 (22)        &       S         &       14.398  $\pm$           0.019   &       -2.119  $\pm$           0.163&3\\

\hline\noalign{\smallskip}

18&     SOON/NOAA\tablefootmark{a}      &       1997 - 2002 (23)        &       N+S         &       14.467  $\pm$           0.019   &       -2.511  $\pm$           0.173&3\\
19&     SOON/NOAA\tablefootmark{a}      &       1997 - 2002 (23)        &       N         &       14.497  $\pm$           0.029   &       -2.259  $\pm$           0.243&3 \\
20&     SOON/NOAA\tablefootmark{a}      &       1997 - 2002 (23)        &       S         &       14.433  $\pm$           0.035   &       -2.719  $\pm$           0.243&3\\

\hline\noalign{\smallskip}

\end{tabular}
\end{center}

\tablefoot{
N denotes the northern hemisphere, S denotes the southern hemisphere, and N+S denotes both hemispheres together. Col. 2: GPR - Greenwich Photoheliographic Results, SOON/NOAA - Solar
Observing Optical Network/United States Air Force/National
Oceanic and Atmospheric Administration, KSO - Kanzelh{\"o}he Observatory for Solar and Environmental Research data. Type of the tracers:
\tablefoottext{a}{sunspot groups}
\tablefoottext{b}{sunspots and sunspot groups}
}

\tablebib{
(1)~\citet{balthvawo1986}; (2) \citet{pulktuo1998}; (3) \citet{javaraiah2003}; (4) \citet{2002SoPh..206..219K}; (5) \citet{lustig1983}}

\end{table*}

%**********************************************************************************************************************
%**************************************************METHODS****************************************
\section{Measurements and data reduction} \label{Measurements and data reduction}

In Paper I we processed KSO sunspot drawings for SCs 20 - 23 (1964 - 2008) and KSO white-light images for SC 24 (2009 - 2016). We used two procedures to determine the heliographic positions of sunspot groups: an interactive procedure (KSO sunspot drawings, 1964 - 2008), and an automatic procedure (KSO white-light images, 2009 - 2016). The KSO observations, data products, and sunspot drawing were described by \citet{2016SoPh..291.3103P,2021SoPh..296..164P}. For the interactive procedure, a software package called Sungrabber\footnote{http://www.zvjezdarnica.hr/sungrabber/sungrabb.html} \citep{2007CEAB...31..273H} was used. The area-weighted centres of sunspot groups were estimated by naked eye, while more importance was given to the more pronounced umbra and penumbra. For the automatic procedure, a morphological image processing of KSO white-light images was used, based on the algorithm by \citet{2011IAUS..273...51W}. The final data set consists of the measured properties of the sunspot groups (size, umbra, penumbra, and position) for every observing day and is available at the KSO ftp server\footnote{http://cesar.kso.ac.at/main/ftp.php} as raw fits images\footnote{ftp://ftp.kso.ac.at/phokaD/FITS/synoptic/} and data files\footnote{ftp://ftp.kso.ac.at/sunspots/drawings/automatic/}.

We used two different methods to determine the synodic angular rotation velocities: a daily shift (DS) method, and a robust linear least-squares fit (rLSQ) method. With the DS method, the synodic rotation velocities were calculated from the daily differences of the central meridian distance $CMD$ and the elapsed time $t$, while with the rLSQ method, they were calculated by fitting a line to the measured positions in time, $CMD(t)$, for each tracer. In the second method, the synodic rotation velocity corresponds to the slope of the fit. We also used a robust fit because the measured data occasionally have outliers due to a false identification and for other reasons. It assumes an iteratively reweighting least-squares fit with Huber's t weighting function \citep{huber1981}. Thereafter, we converted the synodic into sidereal rotation velocities \citep{1995SoPh..159..393R,2002SoPh..206..229B,2014SoPh..289.1471S}. More details about the instrumentation and measurements are available in Paper I (Sect. 2), which also offers more details about the determination of the heliographic positions and rotation velocities (Paper I, Sect. 3 and 4).

Although rotation velocities were determined by two methods (DS and rLSQ), we used the DS method for the statistical analysis in the second part of Sect. \ref{Cycle related variations of the DR} and in Sect. \ref{Long-term variations of the DR} because the DS method yields a higher number of velocities for each bin and is better in filtering out erroneous data points since just two position measurements are taken into account during the velocity calculation. With the rLSQ method, incorrect position measurements are more likely to slip into the final data set because the number of position measurements calculated for each sunspot group is in the range 3 - 11. We concluded in Paper I that to calculate rotational velocities, the DS method is more reliable than the rLSQ method during SCs 20 - 23, while the rLSQ method is more reliable during SC 24. As the DS method is more reliable during a longer period of time (four out of five analysed SCs), this is another reason to prefer it for the further analysis.

The DR parameters $A$ and $B$ for the whole cycles, necessary for the analysis presented in the first part of Sect. \ref{Cycle related variations of the DR}, have already been calculated and are listed in Table 2 of Paper I. For comparison of the obtained rotation profiles, we used a statistical significance similar to that of \citet{1995HvaOB..19....1B}. They regarded the difference between two solar rotation rates as statistically significant if the change in the two rates is larger than the sum of the threefold standard errors of the rotation rates. Otherwise, the difference of the two results (e.g. two parameters $A$: $A_1$ and $A_2$) can be regarded as statistically insignificant, especially when the next criterion is applied,
\begin{equation}
\Delta A = A_1 - A_2 < 1 \sigma(A_1) + 1 \sigma(A_2)
,\end{equation}
where $\sigma$ represents the standard error. Values satisfying Eq. 2 are treated as statistically identical. This statistically insignificant difference of the two results is called ``1 $\sigma$ coincidence" below.

%/////////////////////////////// FIGURE 1 - A_vs_latitude_DS_LSQ ///////////////////////

\begin{figure}[!ht]
   \centering
    \resizebox{8cm}{!}{\includegraphics[bb = 55 25 760 613]{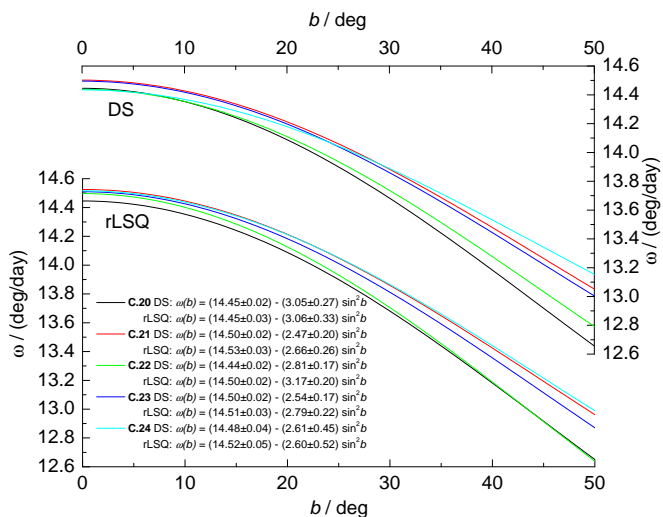}}\
     \caption{DR profiles for five SCs 20 - 24 obtained for both hemispheres together (N+S). The DR profiles were determined using sidereal rotation velocities calculated with the DS method and the rLSQ method. The sidereal rotation velocity is denoted by $\omega$ and the heliographic latitude by $b$.}
     \label{Fig1}
\end{figure}

In the second part of Sect. \ref{Cycle related variations of the DR} and in Sect. \ref{Long-term variations of the DR}, we investigate the relation between solar activity and the DR parameters, $A$ and $B$, for different time periods and binning methods. The rotation velocities were binned according to two different methods, and the DR parameters were determined for each bin by least-square fitting the corresponding DS and rLSQ \mbox{sidereal} rotation velocities for the solar DR law (Eq. 1). In all cases, the two solar hemispheres were treated together to ensure a sufficiently high number of data points also for the periods near minimum of activity. The binning was performed with two methods: 1) each solar cycle was divided into N bins of equal time duration, and then the DR parameters were calculated; 2) the whole observed interval was divided into bins of unequal time duration, but with the same number of observations (rotation velocities).
\newline\newline The first method ensures temporal homogeneity and equal sampling, but produces heteroscedasticity due to the unequal number of observation in each bin. The second method preserves homoscedasticity at the expense of the unequal time duration of the bins. In order to asses the possible impact of the binning on results, we divided each SC into 5, 10, 20, 40, and 50 bins of equal duration, and the whole observed interval into bins with 100, 200, and 500 rotation velocities.

The recalibrated daily, monthly, or yearly mean total sunspot numbers \citep{2014SSRv..186...35C,2015IAUGA..2249591C,2015sac..book...35C}, that is, version 2.0 of the data series provided by the \citet{sidc}, Royal Observatory of Belgium, Brussels, were used as a measure of solar activity for a given time period or binning method. In the statistical analysis (see Table~\ref{Tab2}), the following values were calculated: the slope of the fit with the corresponding standard error (slope $A$ or slope $B$), and Pearson's correlation coefficient $r$ and the corresponding $p$ value, which represents the probability that the null hypothesis (no correlation, $r$ = 0) is accepted. A significance level of 5\% was used. If the $p$ value was lower than the significance level (0.05), the null hypothesis was rejected, suggesting that there is a significant linear relation between rotation and activity. In contrast, if the p-value was higher than the significance level (0.05), there is insufficient evidence to conclude there is a significant linear relation between rotation and activity.

We analysed solar cycle phase variations in the DR parameters (Sect. \ref{Solar cycle phase-related variations of the DR}) and the long-term variations in the DR parameters (Sect. \ref{Long-term variations of the DR}) for the time period 1964 - 2016 (SCs 20 - 24). While analysing the solar cycle phase-related changes in the DR, in addition to the calculated KSO DR and rLSQ parameters, we also used the Debrecen Photoheliographic data (DPD), GPR and USAF/NOAA parameters calculated by \citet{2017SoPh..292..179R}. The USAF/NOAA Sunspot Data and DPD sunspot database \citep{2017MNRAS.465.1259G,2016SoPh..291.3081B} used in \citet{2017SoPh..292..179R} was downloaded from the websites solarscience.msfc.nasa.gov/greenwch.shtm and fenyi.solarobs.csfk.mta.hu/DPD/.
In order to determine the phase of the SC, we used the times of minima and maxima of solar activity provided by the \citet{sidc}, Royal Observatory of Belgium, Brussels\footnote{https://wwwbis.sidc.be/silso/cyclesmm}, and we used an expression similar to the one used in \citet{2017MNRAS.466.4535B}, 
\begin{equation}
\phi = (\tau - M)/\left|(m-M)\right|
,\end{equation}
where $\tau$ is the current time, and $M$ and $m$ are the dates of maximum and minimum of the 11-year cycle closest to $\tau$. When we discuss variations related to odd and even cycles (e.g. in Sect. \ref{Solar cycle phase-related variations of the DR}), which encompass two consecutive 11-year cycles, we calculated the phase in the following way. All points that belong to the ascending branch of the odd cycle are from $\phi = 0$ to $\phi = 0.5$, the points that belong to the descending branch are from $\phi = 0.5$ to $\phi = 1$, and for the even cycle, the ascending branch belongs to the phase range [1, 1.5], while the descending branch belongs to the phase range [1.5, 2].

When we considered the temporal behaviour of two parameters (and not the periodograms), the phase calculated from Eq. 3 was added to the corresponding cycle number such that the time $t=0$ corresponds to the beginning of SC 20, $t=0.5$ to the maximum of SC 20, $t=1$ to the beginning of SC 21, and so on. This transition from time to the phase of the corresponding SC is necessary due to the unequal duration of the SC and the asymmetry inside the cycle.

In order to determine a periodicity in the DR parameters and to support the transition from time to the phase of the SC, we used the standard period-finding periodogram method of \citet{Lomb76} and \citet{Scargle82}  (i.e. Lomb-Scargle periodogram) for unevenly spaced data, and the generalised Lomb-Scargle periodogram \citep{GLS09} to obtain period and statistical significance. Additionally, methods of phase dispersion minimisation \citep{PDM78} and epoch folding \citep{EF83} were also used for the non-sinusoidal analysis. The jackknife method \citep{Jackknife49} was used to estimate the standard error of the obtained period.

The locally estimated scatterplot smoothing (LOESS) method \citep{Cleveland93} was used for the scatterplot smoothing and for the non-parametric regression of DR parameters. For the long-term analysis of DR parameters, we used the moving-average method to remove any short-term in-cycle-related variations. In this method, an average value of a parameter of interest is calculated inside a predetermined time window that corresponds to the detected periodicity.

For the statistical analysis, we used the R and CRAN packages \citep{R-cran}, as well as Python libraries \citep{python}.
In the further text, ``correlation'' means positive correlation, and ``anticorrelation'' means negative correlation.

%**********************************************************************************************************************
%**************************************************RESULTS AND DISCUSSION****************************************

\section{Results and discussion} \label{Results and discussion}
\subsection{Cycle-related variations in the DR} \label{Cycle related variations of the DR}

In this section, the decimal fractions of the years specified in Table 2 of Paper I (SC 20: 1964.8–1976.3, SC 21: 1976.3–1986.7, SC 22: 1986.7–1996.4, SC 23: 1996.4–2008.9, and SC 24: 2008.9–2016.3) were used as start and end points of each SC. They are taken from Braj\v{s}a et al. (2009), except for the minimum in 2008 and the maximum in 2014 (Source: WDC-SILSO, Royal Observatory of Belgium, Brussels\footnote{http://www.sidc.be/silso/news004}).

The DR parameters, calculated separately for SCs 20 - 24 and obtained with the DS and rLSQ methods for both hemispheres together - N+S, the northern hemisphere - N, and the southern hemisphere - S, are available in Table 2 of Paper I (DS: rows 1-5, 8-12, 15-19; rLSQ: rows 22-26, 29-33, 36-40). When we compare the DS and rLSQ measurements, we see that the results for both DR parameters in all cases (northern, southern, or both hemispheres) show a 1 $\sigma$ coincidence. Only in some cases were higher differences observed, but they are still insignificant. This means that the results of the two different methods we used to determine the rotation velocities remain almost identical also for the shorter periods of time (SCs), and not only for the whole time period (1964 - 2016), as we concluded in Paper I.

%/////////////////////////////// TABLE 2 - ROT VS ACT ///////////////////////

\begin{table*}[!ht]
\begin{center}
\caption{Statistical results of the weighted least-square fits of the dependence of the DR parameters, derived by KSO DS data, and solar activity for SCs 20 - 24 (rows 1-5) and for different binning techniques. $N$ is the number of dots; slope $A$ (slope $B$) is the slope of the linear fit; $r$ is Pearson's correlation coefficient; and $p$ is the corresponding $p$-value.}\label{Tab2}
\begin{tabular}
[c]{>{\centering}m{0.5cm}c>{\centering}m{7cm}cccc}\hline\noalign{\smallskip}
\multicolumn{7}{c}{Solar cycles approach (Sect. \ref{Cycle related variations of the DR})} \\\hline\noalign{\smallskip}
Row&Cycle no.&Time period&Slope $A$ &$N$ & $r$& $p$ \\\hline\noalign{\smallskip}

1&20\tablefootmark{b}   &1964.8 - 1976.3\tablefootmark{a}&-0.0012       $\pm$ 0.0005  &12 & -0.59& 0.04       \\
2&21\tablefootmark{b}   &1976.3 - 1986.7\tablefootmark{a}&0.0008        $\pm$ 0.0004  &11&0.49&0.12\\
3&22\tablefootmark{b}   &1986.7 - 1996.4\tablefootmark{a}&0.0001        $\pm$ 0.0004  &11&0.08&0.81\\
4&23\tablefootmark{b}   &1996.4 - 2008.9\tablefootmark{a}&-0.0001       $\pm$ 0.0003  &13&-0.07&0.81\\
5&24\tablefootmark{b}   &2008.9 - 2016.3\tablefootmark{a}&-0.0002       $\pm$ 0.0008  &9&-0.09&0.80\\\hline\noalign{\smallskip}

\multicolumn{7}{c}{Solar cycle phase related approach (Sect. \ref{Solar cycle phase-related variations of the DR})} \\\hline\noalign{\smallskip}
Row&Binning&Time period&Slope $A$ &$N$ & $r$& $p$       \\\hline\noalign{\smallskip}
6&yearly\tablefootmark{b}       &1964-2016&0.0002       $\pm$ 0.0002    &53 & 0.14& 0.31    \\ %cmidrule{3-3}
7&200 velocities/bin\tablefootmark{c}&1964-2016&0.0003  $\pm$ 0.0002    &161&0.11&0.16\\
8&20 bins per cycle\tablefootmark{d}&1964-2016&0.0002   $\pm$ 0.0002    &90&0.12&0.26\\
9&40 bins per cycle\tablefootmark{e}&1964-2016&0.0002   $\pm$ 0.0002    &180&0.10&0.18\\\hline\noalign{\smallskip}

Row&Binning& Time period &Slope $B$  &$N$ & $r$& $p$    \\\hline\noalign{\smallskip}
10 & yearly\tablefootmark{b}            &1964-2016&-0.0012      $\pm$ 0.0019    &40 & -0.11& 0.52\\%cmidrule{3-3}
11&200 velocities/bin\tablefootmark{c}&1964-2016&-0.0026        $\pm$ 0.0018    &161&-0.12&0.13\\
12&20 bins per cycle\tablefootmark{d}&1964-2016&-0.0023 $\pm$ 0.0016    &90&-0.15&0.17\\
13&40 bins per cycle\tablefootmark{e}&1964-2016&-0.0025 $\pm$ 0.0017    &180&-0.11&0.14\\\hline\noalign{\smallskip}
\end{tabular}

\end{center}
\tablefoot{
\tablefoottext{a}{the starting and ending epochs of the corresponding SCs are taken from \citet{2009A&A...496..855B}, except
for the beginning of SC 24, which was taken from SILSO World Data Center (2015), Royal Observatory of Belgium, Brussels \citep{sidc};}\\
\tablefoottext{b}{bin size covers one year}\\
\tablefoottext{c}{bin size covers different time intervals, but includes 200 sidereal velocities by which the DR parameters are calculated}\\
\tablefoottext{d}{20 bins per SC (except for SC 24: 10 bins): bins cover the same parts of the SCs, with a different number of velocities by which the DR parameters are calculated}\\
\tablefoottext{e}{40 bins per SC (except for SC 24: 20 bins): bins cover the same parts of the SCs, with a different number of velocities by which the DR parameters are calculated}\\
}

\end{table*}

The annual DR profiles for SCs 20 - 24  and both hemispheres together (N+S) are shown in Fig.~\ref{Fig1}. The DR parameters collected from different
sources for SCs 20 - 23, using only sunspots and sunspot groups as tracers, are listed in Table~\ref{Tab1}.

We compared the DS and rLSQ measurements derived for SC 20 (Paper I, Table 2, N+S: rows 1 and 22, N: rows 8 and 29, S: rows 15 and 36) with the corresponding values derived by other data sets (Table 1, N+S: rows 1-3, N: rows 4 and 5, S: rows 6 and 7). Our $A$ values are lower by 0.05 deg/day on average (with a maximum difference 0.08 deg/day), in the most cases yielding marginally significant differences (2 $\sigma$ or more). Lower coincidence values may be a consequence of some systematic errors in the KSO data before the 1980s \citep{Balth_Fangme1988,1984SoPh...91...55B,2017A&A...606A..72P}. The same comparison performed for all other SCs 21 - 24 and in all cases (N+S, N, S) yields a 1 $\sigma$ coincidence of the KSO and SOON/NOAA results. Therefore, the KSO data set is well suited for long-term cycle to cycle studies, especially for the years after the 1980s.

The statistical results of the weighted least-square fits of the dependence of the parameter $A$ and solar activity (represented by SILSO relative sunspot numbers as proxies) derived for SCs 20 - 24 are listed in Table~\ref{Tab2}, rows 1-5. Solar cycle 20 shows a 2 $\sigma$ statistically significant negative correlation between the solar equatorial rotation rate and solar activity (slope A = -0.0012$\pm$0.0005, r=-0.59, p=0.04), while the results for SCs 22-24 are not statistically significant and show no correlation (r=0.08, r=-0.07, r=-0.09 respectively). Solar cycle 21 behaves oppositely to SC 20, showing a medium positive correlation (slope A = 0.0008$\pm$0.0004, r=0.49), for which no significance was confirmed at the 95\% confidence level (p=0.12), however.

The lack of a significant linear correlation in SCs 21 - 24 could be explained by a more complex interdependence between DR parameters and solar activity inside the cycle, such as alternation of the anticorrelation-correlation-anticorrelation dependence. This behaviour might impede the detection of any variation and correlation in the cycle if average values are used over the whole cycle. In order to further study the behaviour of the DR parameters, we analysed both in-cycle phase-related (Sect. 3.2.) and global variations in the DR parameters (Sect. 3.3).

\subsection{Variations in the DR related to solar cycle phase} \label{Solar cycle phase-related variations of the DR}

\begin{figure}[!ht]
   \centering
     \resizebox{8.2cm}{!}{\includegraphics[bb = -8 15 500 973]{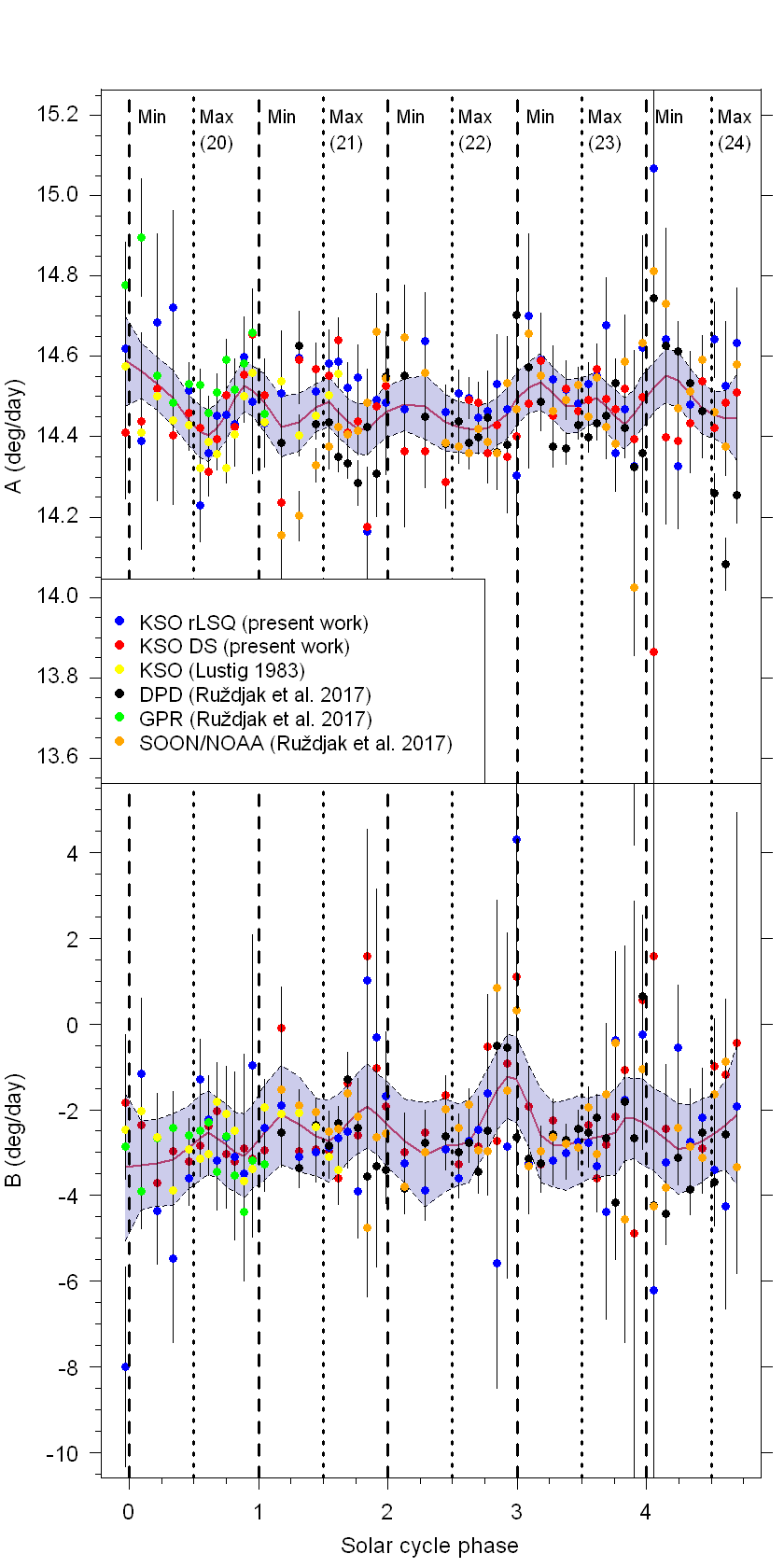}}\
     \caption{In-cycle variation of the equatorial rotation velocity, represented by parameter $A$, and steepness of the differential rotation, represented by $B$, shown in phase space for SCs 20 – 24 (1964 - 2016) and both hemispheres together. Blue and red dots represent our KSO rLSQ and KSO DS values (1964 - 2016), yellow dots show KSO values from \citet{lustig1983} (1964 - 1981), and black, green, and orange dots show DPD, GPR, and SOON/NOAA values from \citet{2017SoPh..292..179R}. The red line represents the result of the LOESS, with the greyish part indicating the 95\% confidence level. Full phases (0-5) correspond to the times of subsequent minima.}
     \label{Fig2}
\end{figure}

In this section we study the in-cycle variations in the DR parameters.
It is well known that the duration of the SC can vary considerably. In order to account for this phenomenon, we made an analysis in the phase space of the SC instead of analysing the temporal behaviour of the DR parameters in the regular time domain. This also took variations in the duration of time intervals between minima and maxima into account. In Fig.~\ref{Fig2}, where the in-cycle variations in the DR parameters $A$ and $B$ are shown, the time represented as the phase of the corresponding SC was obtained from the SILSO times of minima and maxima\footnote{https://wwwbis.sidc.be/silso/cyclesminmax} and applying Eq. 3. Full phase (0-5) corresponds to the time of succeeding minima. Rotational velocities calculated from the position data of four different data sets (DPD, GPR, KSO, and SOON/NOAA) were used to determine yearly $A$ and $B$ values that were then transformed into the phase space. Then LOESS was applied on all available dots (values). Therefore, the dots in Fig.~\ref{Fig2} are values of the DR parameters calculated using the KSO DS and rLSQ angular velocities for the whole time period (1964 - 2016), GPR angular velocities (for the years until 1976), KSO angular velocities from \citet{lustig1983} (for the period 1964 - 1981), and DPD and SOON/NOAA angular velocities (for the years after 1976). We choose DPD, GPR, and SOON/NOAA data sets mainly because the synodic angular velocity differences are smallest for KSO - GPR, KSO - DPD, and KSO - SOON/NOAA combinations when six different data sets are compared; see Fig. 6 in \citet{2011CEAB...35...59P}.

The LOESS curve in Fig.~\ref{Fig2} shows that during the even SCs 20, 22 and 24 $A$ reaches higher values around activity minima and lower values around activity maxima, creating a concave structure. The same finding was already reported \citep{2006SoPh..237..365B,2011A&A...534A..17J,2014SoPh..289..759L,2017SoPh..292..179R}, but, in general, for all analysed SCs, including odd ones. Here, the concave structure is slightly disturbed for odd SCs 21 and 23, with a bump of $A$ (the higher values of $A$) around activity maxima. As appropriate binning of the velocity data is a prerequisite for the analysis of the DR parameters, we repeated the calculation using another binning technique of rotation velocity values and different bin lengths.

\begin{figure}[h]
  \centering
  \resizebox{10.2cm}{!}{\includegraphics[bb = 5 15 560 550]{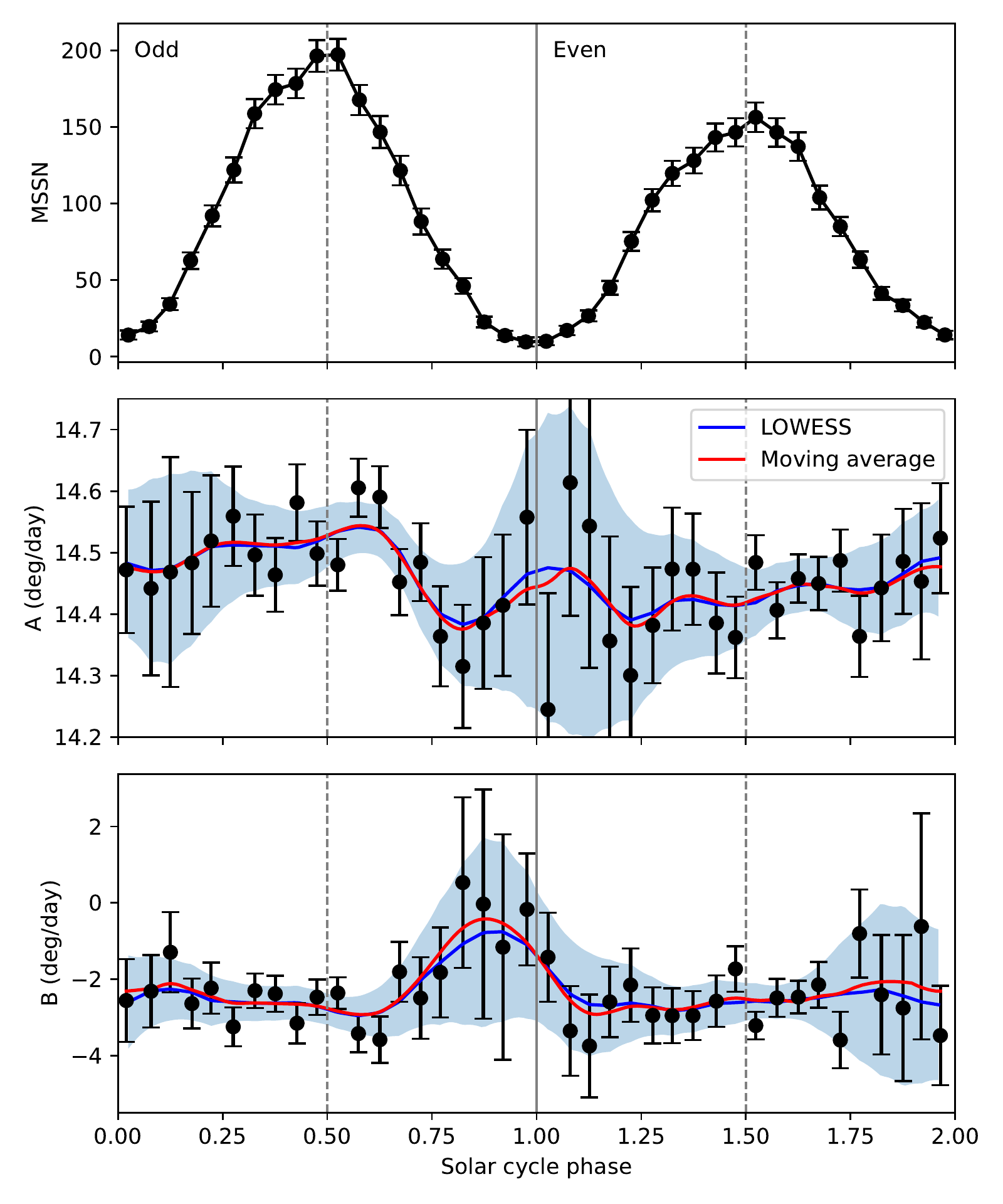}}\
   \caption{Solar cycle phase diagram in the single odd-even space for parameter $A$ (middle panel) and parameter $B$ (bottom panel). For 40 bins of the same size (0.05 phase each), the DR parameters were calculated and are shown as black dots with the corresponding standard errors. A moving average of these values was calculated (red curve), using a window size of 3 bins. Additionally, LOWESS smoothing was applied to the original bins using a data fraction of 1/6 (blue curve). In the top panel, the SILSO 13-month smoothed monthly total sunspot number is shown. Phases in the range 0-1 represent odd SCs 21 and 23, and phases in the range 1-2 show even SCs 20, 22, and 24. Whole phases mark cycle minima (vertical solid grey line), and cycle maxima are located at 0.5 and 1.5 phase values (vertical dashed grey lines). Bootstrapping was applied to determine the uncertainty (blue areas).}
   \label{Fig3}
\end{figure}

To do this, we used different approaches to rotation velocity data binning in addition to the already discussed yearly binning: a) we binned in time intervals with constant duration, and b) we binned in time intervals with a constant number of observations. Binning in intervals of constant duration results in temporal homogeneity, with the possibility of applying different statistical methods that require equal temporal spacing, such as autoregression, cross-correlation, or fast Fourier transforms. Unfortunately, such a binning leads to heteroscedasticity and unequal uncertainties: intervals around maximum activity cover a larger number of sunspots, and therefore the rotational velocity can be obtained with smaller uncertainty. Conversely, during SC minima, only a few sunspots are present, which can result in high uncertainties in the velocity determination. Homoscedasticity and similar uncertainties are preserved when the bins are made of the same number of observations, but in this case, the temporal homogeneity is lost: bins at solar minima cover a longer time period than those at solar maxima. Therefore, many more measurements of the DR parameters are determined around solar maxima than at minima. As there are advantages and disadvantages for both binning techniques, we performed both, but show only the results for time intervals with constant duration (Fig.~\ref{Fig3}) as a similar behaviour was obtained in both cases.

%/////////////////////////////// FIGURE 6 - B_KSO_DPD_SOON ///////////////////////

A periodical pattern of concave structure is noticeable in Fig. 2. As the sunspot number does not show a monotonously increasing or decreasing behaviour in time, but rather a quasi-periodic behaviour, we cannot expect to find a simple correlation between the DR parameters, as is shown by the lack of significant linear regression in each of the analysed SCs (Table 2, rows 1-5). An interdependence and possible correlation between solar rotation (DR parameters) and solar activity (sunspot number) can still be established if the same or a similar periodicity for both the DR parameters and sunspot numbers is found. Therefore, we used a standard Lomb-Scargle period-finding technique as well as a generalised Lomb-Scargle method that can include the uncertainties of the DR parameters. Parameter $A$ shows an average periodicity of $1.89 \pm 0.10$ SC when the weighted generalised Lomb-Scargle method is used, with a periodicity of $1.86 \pm 0.11$ SC and $1.96 \pm 0.09$ SC for a binning with 200 velocities/bin and 40 bins/cycle. A high statistical significance was obtained, with $p$-values between $1.2\cdot10^{-4}$ to $5.8\cdot10^{-6}$ for different bin lengths and methods, which means that the probability for a noise to be detected as a periodic signal is between $1:8000$ to $1:170\,000$. These results were confirmed by methods of epoch folding and phase dispersion minimisation (PDM).

To validate the procedure, we also obtained an average period of $1.02 \pm 0.02$ SC for SILSO sunspot numbers, and $1.02 \pm 0.02$ SC and $1.01 \pm 0.01$ SC for a binning with 200 velocities/bin and 40 bins/cycle. The period is in accordance with the expected value. For parameter $B$, an average period of $0.92 \pm 0.06$ SC was found with the weighted generalised Lomb-Scargle method. Again, all of the periods were confirmed by methods of epoch folding and PDM.

According to these results, parameter $A$ is clearly correlated with the full 22-year SC, that is, with two SCs, while parameter $B$ seems more correlated with one SC. Therefore, in order to analyse in-cycle phase-related variations in the DR parameters, periodograms with a period of two SCs were constructed (Fig.~\ref{Fig3}).

Fig.~\ref{Fig3} shows a SC phase diagram for the angular sidereal rotation rate of the equator, represented by parameter $A$ (middle panel), and the steepness of the DR, represented by parameter $B$ (bottom panel), along with the SILSO 13-month smoothed monthly total sunspot number shown in the upper panel. A single odd-even cycle system in phase space was used, as in Figures 2 and 3 of \citet{2017MNRAS.466.4535B}, with a corresponding phase range of [0,2]. This range was divided into 40 bins with a phase width of 0.05. For each bin, parameters $A$ and $B$ were calculated from sidereal rotation velocities and are shown as black dots. Then, a moving average was performed using a window size of 3 bins (red curve), and LOWESS smoothing was applied to the original bins using a data fraction of 1/6 (blue curve). The whole KSO DS data set (1964-2016) was used. The odd part of the figure covered SCs 21 and 23, and the even part covered SCs 20, 22, and 24.

Fig.~\ref{Fig3} shows that $A$ is lowest in the minimum epochs between the odd and even cycles (just before phase 1.0) and reaches higher values around the odd minima (around phases 0 and 2.0). The main difference to Figure 2 from \citet{2017MNRAS.466.4535B}, which states a monotonous decrease of $A$ during the whole odd cycle and a monotonous increase of $A$ during the whole even cycle, is the occurrence of the bump for $A$ around the odd activity maximum (phase 0.5).

A negative correlation of $A$ and the solar activity, which implies higher values of $A$ during the minimum of the magnetic activity, and vice versa, lower values of $A$ during the maximum of the magnetic activity, is predicted by theoretical models \citep{2004SoPh..220..333B,2004ApJ...614.1073B,2006MNRAS.373..819L,2007A&A...471.1011L,2014IAUS..302..114B,2016AdSpR..58.1507V} and has been found in a number of observational studies \citep{lustig1983,gilmanhow1984,balthvawo1986,2006SoPh..237..365B,2011A&A...534A..17J,2014SoPh..289..759L,2017SoPh..292..179R}. This behaviour is also confirmed here for in-cycle variations in Fig.~\ref{Fig2}, with disturbances, the  bumps, during the odd maxima. In odd-even phase space (middle part of Fig.~\ref{Fig3}), this behaviour, which can be described by a concave structure, is weak but recognizable for even cycles. During the odd cycles (middle left part of Fig.~\ref{Fig3}), a bump is again observed in the middle. Therefore, the bump in $A$ is observed (using our KSO data) regardless of the binning method. It is not clear why it occurs, whether it is the result of a systematic error or indeed the behaviour of $A$ during the odd cycles of the analysed time period.

Fig.~\ref{Fig2} and Fig.~\ref{Fig3} also show that $B$ becomes more negative, that is, the Sun rotates more differentially, around SC maxima (Fig.~\ref{Fig2} around phases 1.5, 2.5, 3.5, and 4.5; Fig.~\ref{Fig3} around phases 0.5 and 1.5). These flat parts are characterised by $B$ values around -3 deg/day, similar to the values derived by \citet{2017MNRAS.466.4535B}. In the descending part of the cycles (phases: 0.5-1, 1.5-2.0), $B$ becomes less negative, that is, the Sun rotates more rigidly. This is in contrast to theoretical calculations, which suggest a more rigid rotation during cycle maxima.

\begin{figure}[!ht]
 \centering
        \subfloat{\label{Fig2a}\resizebox{7cm}{!}{\includegraphics[viewport=130 125 700 550]{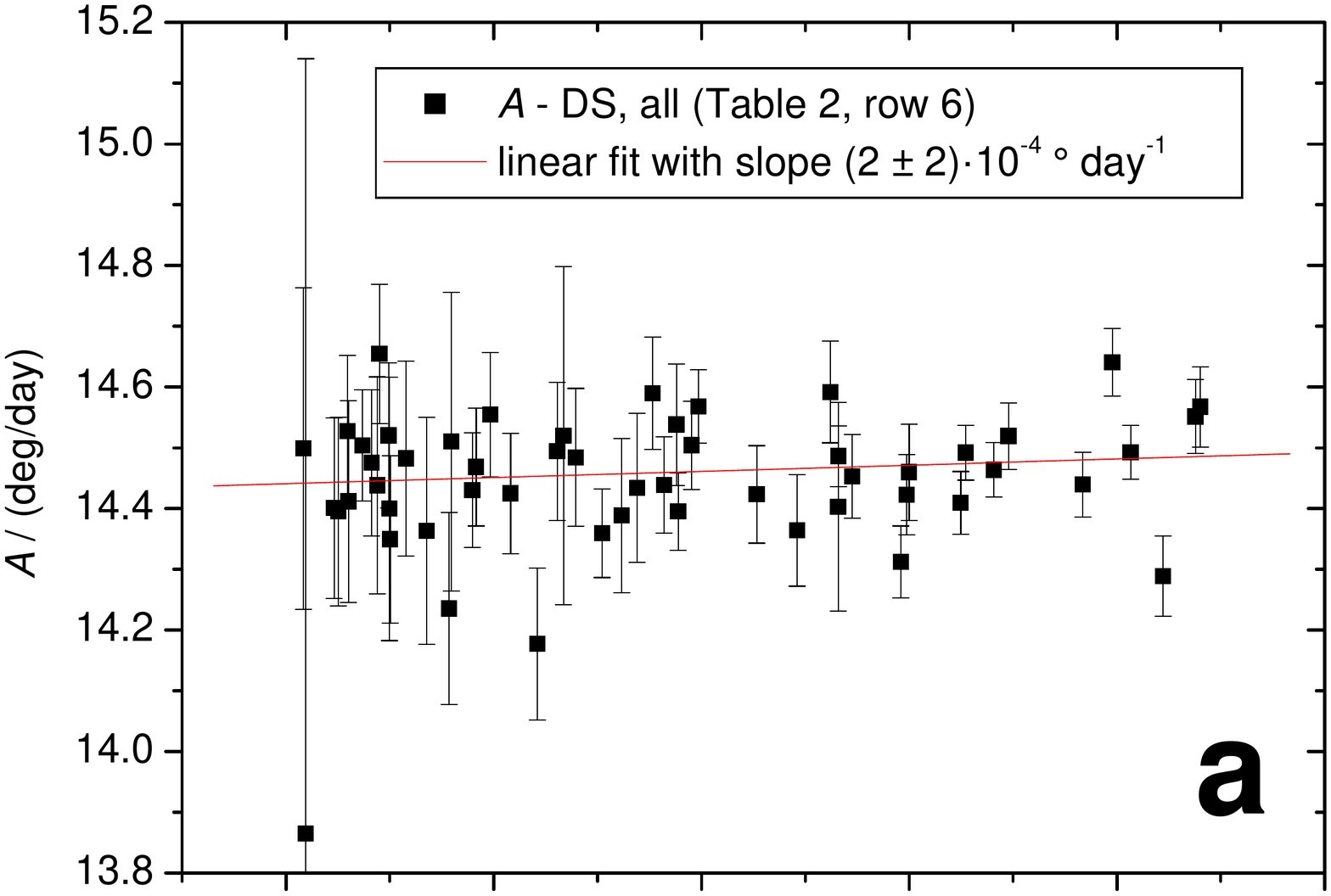}}}\\
        \subfloat{\label{Fig2aa}\resizebox{7cm}{!}{\includegraphics[viewport=130 85 700 520]{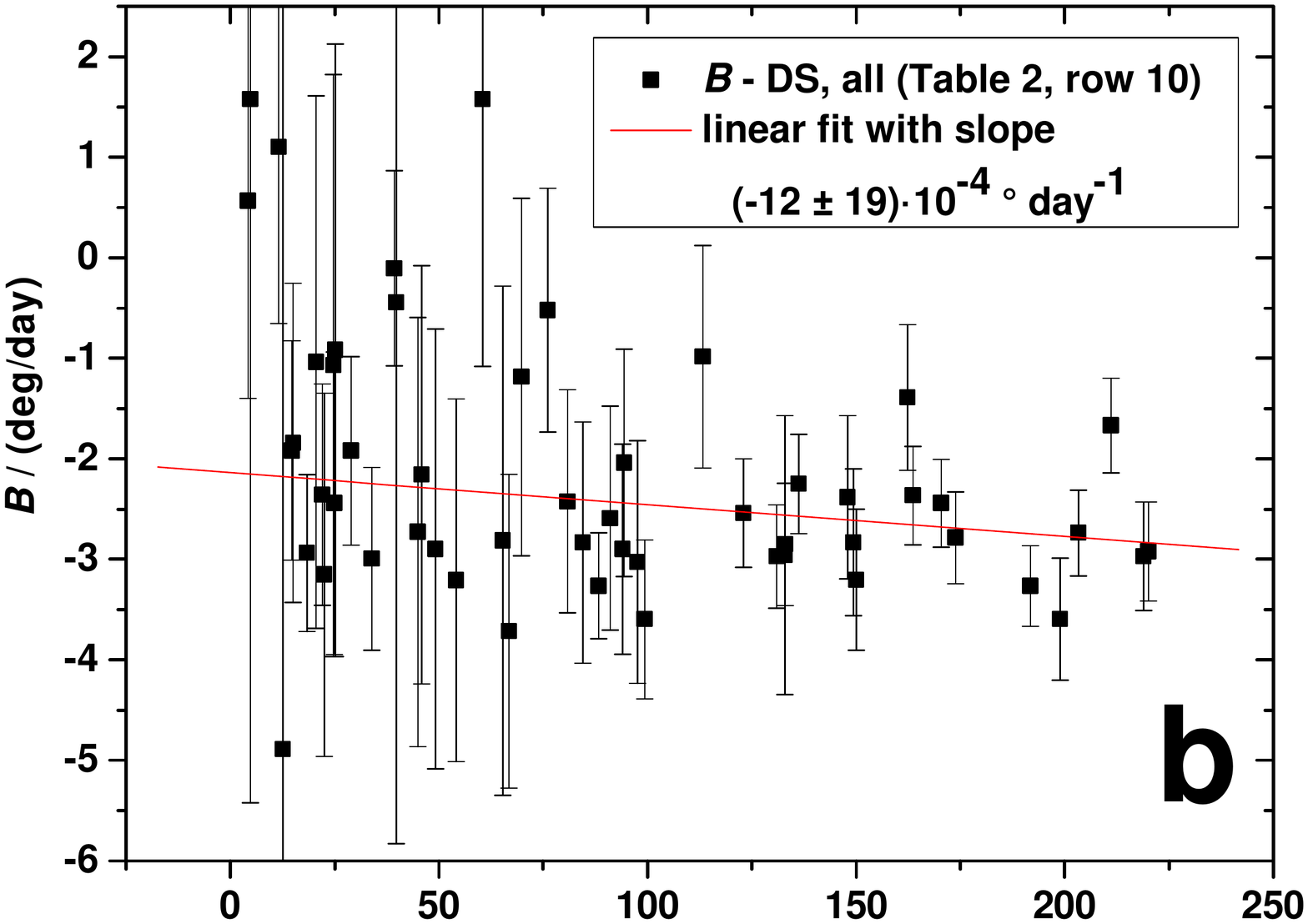}}}\\
\caption{Dependence of annual KSO DR parameters $A$ and $B$ and the WDC-SILSO relative sunspot number as an indicator of solar activity. Straight lines represent least-squares fits through the set of data points derived by the DS method for individual years. Parts a and b cover all KSO measurements 1964 - 2016. Statistical results for each least-squares fit are available in Table~\ref{Tab2}, rows 6 and 10.}\label{Fig4}
\end{figure}

Although \citet{2017MNRAS.466.4535B} did not use tracers in the photosphere, but coronal green emission lines from the corona and the time periods do not coincide (photosphere: 1964-2016, corona: 1939–2001), almost the same behavioural pattern of $A$ and $B$ is observed. The coronal DR was discussed in many papers that studied line emissions in the corona or tracers like coronal bright points \citep{1997SoPh..171....1B,2010NewA...15..135B,2011A&A...534A..17J,2017MNRAS.466.4535B}, and it was concluded that the coronal DR agrees well with the photospheric DR \citep{2004A&A...414..707B,2010A&A...520A..29W,2015A&A...575A..63S}. Therefore, our paper confirms and reinforces the anticipated photosphere-corona differential rotation coupling. Our results for the DR parameters obtained in the single odd-even cycle phase system (Fig.~\ref{Fig3}) are also comparable to the one obtained in the phase space of separate SCs (Fig.~\ref{Fig2}), even though in the latter, multiple data sets have been used.

\subsection{Long-term variations in the DR} \label{Long-term variations of the DR}

In this section we present our study of the long-term changes in the DR parameters with the aim to relate them with the solar activity on Gleissberg-cycle scales \citep{2003JGRA..108.1003P,2017LRSP...14....3U,2020LRSP...17....2P}. The statistical results of the weighted least-squares fits of the dependence of the DR parameters $A$ and $B$ and solar activity, represented by relative sunspot numbers as proxies, were derived for different binning techniques: a yearly binning during 1964 - 2016, an equal number of bins per analysed SC and a different number of bins per SC (but with the same number of sidereal rotation velocities in each bin). The results are listed in Table~\ref{Tab2}: rows 6-9 for $A,$ and rows 10-13 for $B$.

\begin{figure}[!ht]
 \centering
        \subfloat{\label{Fig2a}\resizebox{4.5cm}{!}{\includegraphics[viewport=125 55 360 410]{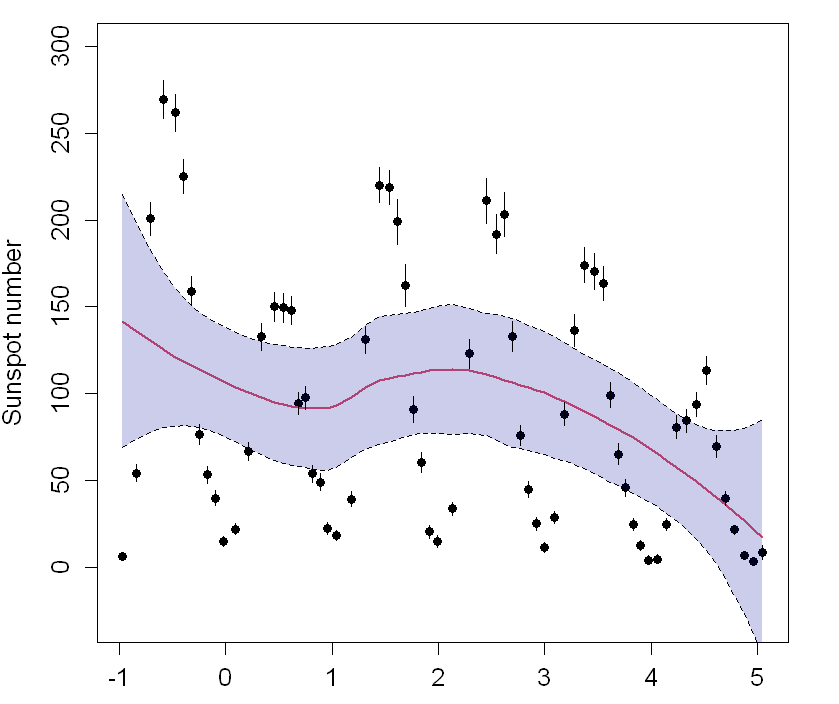}}}\\
        \subfloat{\label{Fig2aa}\resizebox{4.5cm}{!}{\includegraphics[viewport=-135 -55 445 480]{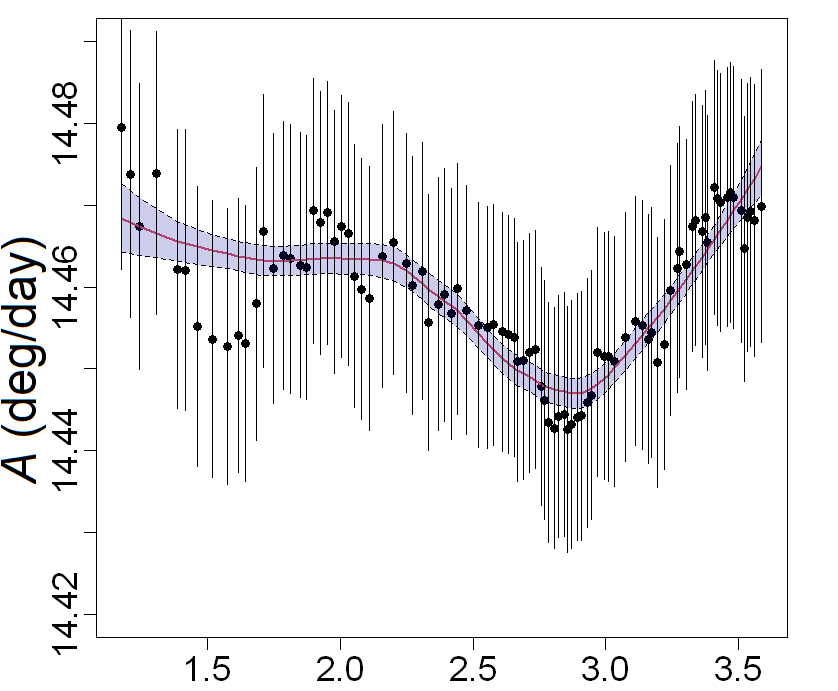}}}\\
        \subfloat{\label{Fig2aaa}\resizebox{4.5cm}{!}{\includegraphics[viewport=-163 25 555 440]{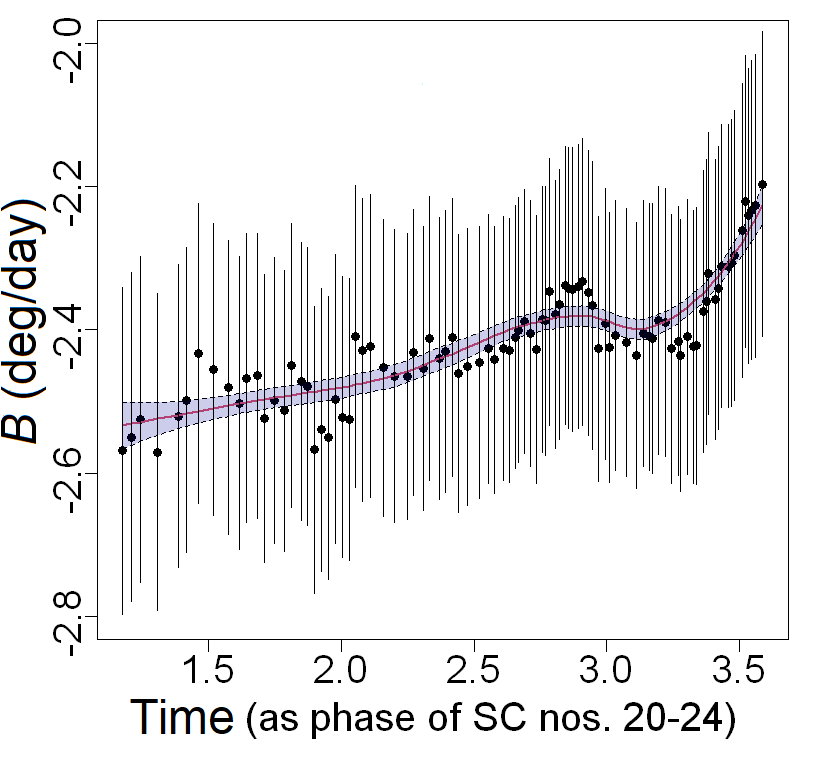}}}\\\caption{Long-term changes in the KSO differential rotation parameters. In the top panel SILSO sunspot group numbers are shown in phase space, where the whole phases mark cycle minima. Phase range 0-1 belongs to SC 20, phase range 1-2 to SC 21, etc. Phase 0 indicates the beginning of the analysed time period (1964). A LOESS non-parametric fit was applied to the original data using span=0.65, and a smoothing in a window covering 65\% of data points (red curve) in order to obtain general trend. The same was performed for parameters $A$ and $B$ (middle and bottom panel), but with moving-average values in order to remove in-cycle variations, with a LOESS smoothing of the same span as before. Moving averages were calculated over two solar cycles for each of the data points, which corresponds to the periodicity of parameter $A$.}\label{Fig5}
\end{figure}

According to the results for $A$ (Table~\ref{Tab2}, rows 6-9), a non-significant positive correlation is obtained in all cases, which is in contrast to theoretical calculations. Annual values (Table~\ref{Tab2}, row 6) are displayed in panel (a) of Fig.~\ref{Fig4} for illustration. Straight lines represent least-squares fits through the set of data points derived by the DS method for individual years. Because the calculation of the slope of the linear fit is sensitive to the standard error of individual values of the DR parameters, a weighted least-squares fit was applied, thus giving more importance to measurements with smaller error bars.  According to the theoretical model, a negative correlation is expected for $A$, that is, a decreasing linear function, which was not obtained here.

We assume that our statistical results of the weighted least-squares fits of the dependence of parameter $A$ and solar activity (for the KSO data set, 1964-2016) does not yield the expected negative correlation due to the in-cycle variations that show a complex correlation behaviour, and because this time span covers the years just after the modern Gleissberg maximum, which includes an alternation of the correlation and anticorrelation dependence of parameter $A$ and solar activity, as explained below.

A secular increase in the magnetic activity during most of the twentieth century has been confirmed, reaching its maximum in the second half of the twentieth century, that is, around 1970 \citep{2003JGRA..108.1003P,2017LRSP...14....3U,2020LRSP...17....2P}.  Since then, magnetic activity has been decreasing, which is shown in the top panel of Fig. 5, where long-term changes in solar activity represented by the SILSO sunspot numbers are shown. Moreover, \citet{2004HvaOB..28...55B}, \citet{2006SoPh..237..365B}, \citet{2014SoPh..289..759L}, \citet{2017SoPh..292..179R} and \citet{2020LRSP...17....2P} have confirmed the secular decrease in the rotation during the twentieth century, with an equatorial rotation velocity flip, that is, a beginning speed-up, being reported for the 1990s \citep{2015A&A...575L...2Z,2017SoPh..292..179R}.

In order to remove in-cycle variations and obtain only general trends, we calculated the moving averages of DR parameters $A$ i $B$ by averaging each data point over two SCs (parameter periodicities as obtained in Sect. 3.2). The middle panel of Fig. 5 shows the results for $A$ . The rotation flip, that is, the beginning speed-up, occurs around phase $2.8$ ($2.84 \pm 0.03$), that is, in the early 1990s ($1994.5 \pm 0.5$). Therefore, it is obvious that the activity flip (maximum of SC 19) and the rotation flip do not occur simultaneously, but with a certain phase shift.

The phase shift approximately coincides with SCs 20, 21, and 22, during which solar activity represented by the sunspot number and the equatorial rotation velocity represented by $A$ both show a secular decreasing trend, indicating a possible correlation of the two quantities. Ultimately, this might be related to the positive signs in the slope of $A$ and in the correlation coefficient $r$ for SCs 21 and 22 (Table~\ref{Tab2}, rows 2 and 3). An increasing trend for $A$ after phase 2.8 accompanied with decreasing trend for the sunspot number yields an anticorrelation of the two quantities, and may be related to the negative signs in the slope of $A$ and to the correlation coefficient $r$ for SCs 23 and 24 (Table~\ref{Tab2}, rows 4 and 5). According to these results, during the analysed time period of KSO data (1964-2016), an alternation of the correlation and anticorrelation dependence of the parameter $A$ and solar activity is present. This violates the anticorrelation of the two quantities for the whole period that is generally predicted by the theoretical model.

Therefore, the suggestion is that when the relation of the rotation and activity with the data sets that include the Gleissberg maximum or minimum are analysed, it is necessary to analyse separately parts of the data set in between the different flips (e.g. before the activity flip, between the activity and rotation flip, and after the rotation flip). In other words, the final result depends on the time period that is included in the analysis, that is, it does matter which part of the Gleissberg cycle we consider.

For instance, the \citet{2017SoPh..292..179R} results for $A$ versus activity show a statistically insignificant negative correlation for the period 1874-1976 (GPR) and a statistically significant negative correlation for the period 1977-2016 (USAF/NOAA data). When the sets are combined, a significant negative trend is obtained for the whole period 1874-2016. We therefore again conclude that the final result depends on the time period that is included in the analysis. As a non-significant negative correlation is obtained for the time period 1874-1976 (GPR), something clearly decreases the significance of the negative correlation in this period, very possibly the correlation of the sunspot number and $A,$ which appears just after the Gleissberg maximum (around  the 1970s) and ends with the equatorial velocity flip (in the 1990s). We also assume that including the minimum of the Gleissberg cycle (around the 1900s) in the GPR data set, by which the disputed insignificant negative correlation was derived, has the same effect on decreasing the significance of the negative correlation. Further analysis using the GPR data, as well as a statistical method that can take care of the phase shift between Gleissberg maximum and the rotation flip, is necessary to confirm this assumption.

 \citet{2011A&A...534A..17J} analysed the same relation ($A$ versus relative sunspot number, e.g. activity) for the time span 1998-2006. It covers the period after the early 1990s, that is, after the equatorial velocity flip, and yields a significant negative correlation. Although this result is based on the analysis of small bright coronal structures, it agrees with our result of a negative correlation between $ A $ and activity after phase 2.8 (after the 1990s).

The results for DR parameter $B$ are given in Table~\ref{Tab2} (rows 10-13) and for the yearly values displayed in the lower panel of Fig.~\ref{Fig4} (part b). We obtained anticorrelations (insignificant) for the yearly values as well as for the binned values, which is again in contrast to the theoretical predictions \citep{2004ApJ...614.1073B,2006MNRAS.373..819L,2007A&A...471.1011L,2014IAUS..302..114B,2016AdSpR..58.1507V}. They expect a correlation of parameter $B$ and solar activity, which means a more pronounced differential rotation (more negative value of $B$) during the minimum of magnetic activity. This actually means that the Sun should rotate more differentially during activity minimum and more rigidly during activity maximum.

The bottom panel of Fig. 5 shows long-term changes in $B$. In general, $B$ increases, that is, it becomes less negative. If moving averages of parameter $B$ and the sunspot number are used, averaged over two SCs (as presented in Fig. 5, lowest panel), a statistically highly significant anticorrelation between $B$ and SN is found with r$=-0.67$ and p$=2\cdot10^{-13}$, with a slope of $(-0.0044\pm0.0005)$ deg/day. This means that we observe an anticorrelation of $B$ and activity (more rigid rotation when the activity decreases) in the analysed KSO period 1964 - 2016, which is in accordance with Table 2 (rows 10-13) and is opposite to the theoretical model. However, it is worth pointing out that around phase 3.0, a weak correlation of $B$ and activity is observed.

\citet{2017SoPh..292..179R} pointed out that for the yearly parameter values, the sign of correlation for $B$ changes. In the time period 1874-1976 (GPR data), it is insignificantly negative, and in the time period 1977-2016 (USAF/NOAA and DPD data), it is significantly positive. When the sets are combined, the larger GPR data set prevails and an insignificant negative trend is obtained for the whole period 1874-2016. Likewise, our (insignificant) negative correlation obtained with yearly $B$ values for the KSO period 1964-2016 (Table 2, rows 10-13) is very likely a consequence of combining the two trends, weak positive (Fig. 5, bottom panel, around phase 3.0) and strong negative (other phases). It is also very likely a consequence of in-cycle oscillations in the parameter values, which are not excluded when yearly values are used.

 \citet{2011A&A...534A..17J} analysed the same relation ($B$ versus relative sunspot number, i.e. activity) for the time span 1998-2006, that is, for SC 23. An insignificant negative correlation of $B$ and activity was obtained, which is in accordance with the anticorrelation during this period we observed here (see Fig. 5, top and bottom panels, phases 3.0 - 4.0). The most probable explanation for the disagreement of these experimental results with theoretical studies may be the fact that we analysed the period immediately after the Gleissberg maximum, during which non-simultaneous activity and an equatorial rotation velocity flip occur. Perhaps this is also being reflected in the behaviour of parameter $B$. Therefore, it is necessary to repeat the same analysis using data sets that cover a longer period of time that encompasses one Gleissberg cycle or more, depending on availability.

%******************************************************************************************
%***********************************CONCLUSIONS ******************************************
\section{Conclusions} \label{Conclusions}

We analysed the variation in the solar differential rotation and activity using 53 years of KSO data (1964 - 2016), which cover five SCs 20-24. For the first time,  the temporal variation of the DR, as well as the relation between the rotation and activity, were analysed based on KSO data for the years after 1980s.

Concerning the solar cycle-related variations of the DR, the results of the two different methods, DS and rLSQ, used to determine the rotation velocities are almost identical, that is, a statistically significant coincidence was observed for both the longer period (whole data set covering several SCs, Paper I) and for shorter periods (individual SCs). The comparison of the KSO DR parameters and parameters collected from other sources (e.g. SOON/NOAA) for SCs 21 - 24 again yielded a statistically significant coincidence. Therefore, the KSO data set used in this paper is well suited for long-term cycle to cycle studies, especially for the years after 1980s.

The main conclusions for the solar cycle phase changes in the DR parameters are listed here. For even SCs 20, 22, and 24 during the neighbouring minima, $A$ increases, while it drops towards the maximum in between, creating a concave structure. For odd SCs 21 and 23, the concave structure is slightly disturbed with a bump (the higher values) in $A$ around activity maxima. The most pronounced DR (less rigid rotation) is observed around SC maxima, while the most rigid rotation is observed one or two years before each minimum.

Parameters $A$ and $B$ both show periodic cycle-related variations. The periodicity of parameter $A$ is correlated with the two SCs, while parameter $B$ seems to have a periodicity of one SC. Therefore, the obtained periodicities of these parameters, which correspond to the periodicity of the sunspot number and solar activity, clearly show a correlation with the SC.

Evidence for the changes in parameter $A$ has been reported by \citet{lustig1983}, \citet{gilmanhow1984}, \citet{balthvawo1986}, \citet{2006SoPh..237..365B}, \citet{2011A&A...534A..17J}, \citet{2014SoPh..289..759L}, \citet{2017SoPh..292..179R}, and \citet{2017MNRAS.466.4535B}, confirming that solar cycle phase changes in parameter $A$ are a consequence of the suppression caused by strong magnetic fields during magnetic activity maximum. This (reduced equatorial velocity, i.e. lower values of $A$ during the strong magnetic fields of the 11-year Schwabe cycle) is confirmed with the present analysis. In-cycle changes of the parameter $B$ are again in contrast to theoretical predictions \citep{2004ApJ...614.1073B,2006MNRAS.373..819L,2007A&A...471.1011L,2014IAUS..302..114B,2016AdSpR..58.1507V} as we observe more negative values of parameter $B$, that is, a more pronounced differential rotation during activity maximum. The comparison of the present paper and  \citet{2017MNRAS.466.4535B} solar cycle phase diagrams in the single odd-even space confirmed and reinforced the previously stated agreement of photospheric and coronal differential rotation.

The main result of our analysis arises from the analysis of the long-term changes in the differential rotation parameters. We found a phase shift between the activity flip (modern Gleissberg maximum in the 1970s) and equatorial rotation velocity flip (begin of the speed-up in the early 1990s), which coincides with SCs 20, 21, and 22. During this time period, solar activity, represented by the sunspot number, and equatorial rotation velocity, represented by $A$, both show a secular decreasing trend, indicating a correlation of the two quantities. In contrast, after this period, an anticorrelation of the two quantities is confirmed. Therefore, the combination of the correlation and anticorrelation behaviour in the inspected time period can explain the statistical results of the insignificant (positive) correlation of the dependence of DR parameter $A$ and the solar activity for the whole data set (KSO, 1964 – 2016).

It seems that including the period between the Gleissberg maximum (time at which the activity flip occurs, around the 1970s) and time at which the rotation flip occurs (around 1990s) in our data set violates the negative correlation predicted from theoretical studies. In other words, our result suggests that a positive correlation could be a property of the aforementioned phase-shift periods, which occur around the maximum of the Gleissberg cycle, but very likely also around the minimum of the Gleissberg cycle (which should be confirmed with a data set that covers a longer period of time). Therefore, our suggestion is that when the relation of the rotation and activity is analysed, either theoretically or experimentally, parts of the data set in between different flips should be analysed separately (e.g. before the activity flip, between the activity and rotation flip, and after the rotation flip).

The long-term variation  in parameter $B$ shows a secular increase in $B$ ($B$ becoming less negative) during the analysed time period. This implies on long-term scales an anticorrelation of $B$ and activity (more rigid rotation when the activity decreases), which is in contrast to theoretical modelling predictions. There is an indication that the correlation is present during the end of SC 22 and the beginning of SC 23, however. Therefore, again, due to combination of correlation and anticorrelation behaviour in the inspected time period, it is not unexpected that the statistical results of the dependence of the DR parameter $B$ and solar activity for the whole data set (KSO, 1964 – 2016) show an insignificant (negative) correlation and statistically significant anticorrelation when in-cycle variations are removed by the use of moving averages over the whole SC. The most probable explanation for the disagreement of these observational results with theoretical studies may be the fact that we analysed the period immediately after the Gleissberg maximum, during which non-simultaneous activity and equatorial rotation velocity flip occurs. Perhaps this is also being reflected in the behaviour of parameter $B$. Therefore, it is necessary to repeat the same analysis using data sets that cover a longer period of time that encompasses one Gleissberg cycle or more, depending on availability.

\begin{acknowledgements}

We acknowledge the staff of the Kanzelh{\"o}he Solar Observatory (Austria), Royal Observatory of Belgium (Brussels) and the staff of the Heliophysical Observatory (Debrecen, Hungary) for maintaining and organizing the KSO, WDC-SILSO and DPD databases, respectively. This work was supported in part by Croatian Science Foundation under the project 7549 \textquotedblleft Milimeter and submilimeter observations of the solar chromosphere with ALMA\textquotedblright\ and in part by the University of Rijeka under the project uniri-prirod-18-3-1129. It has also received funding from the Horizon 2020 project SOLARNET (824135, 2019–2022). We also acknowledge the support from the Austrian-Croatian Bilateral Scientific Projects "Comparison of ALMA observations with MHD-simulations of coronal waves interacting with coronal holes" and "Multi-Wavelength Analysis of Solar Rotation Profile". IPB thanks the Kanzelh{\"o}he Solar Observatory for the hospitality during her stay at Kanzelh{\"o}he.

\end{acknowledgements}

\bibliographystyle{aa.bst}
\bibliography{Ref_Kanz}

\end{document}